\documentclass[]{aa}
\usepackage{graphicx,url,twoopt,natbib}
\usepackage[varg]{txfonts}
\usepackage{subfigure} 
\usepackage{enumerate}

\bibpunct{(}{)}{;}{a}{}{,} 
\usepackage[colorlinks,
            linkcolor=red,
            anchorcolor=blue,
            citecolor=blue]{hyperref}

\begin{document}

\title{The statistical properties of early-type stars from LAMOST DR8}
\author{{Yanjun Guo}
          \inst{1,2,3},
          Chao Liu\inst{4,2},
          Luqian Wang\inst{1},
          Jinliang Wang\inst{1},
          Bo Zhang\inst{4},
          Kaifan Ji\inst{1},
          ZhanWen Han\inst{1,2,5},
          \and
          XueFei Chen\inst{1,2,5}
          }
\authorrunning{Y. Guo
          \inst{1,2,3},
          C. Liu\inst{4,2},
          L. Wang\inst{1},
          J. Wang\inst{1,2,3},
          B. Zhang\inst{4}
          K. Ji\inst{1}
          Z. Han\inst{1,2,5}
          \and
          X. Chen\inst{1,2,5}
          }
\institute{
   $^1$Yunnan observatories, Chinese Academy of Sciences, P.O. Box 110, Kunming, 650011, China; cxf@ynao.ac.cn\\
           $^2$ School of Astronomy and Space Science, University of Chinese Academy of Sciences, Beijing, 100049, People's Republic of China; liuchao@bao.ac.cn\\
            $^3$ Key Laboratory for Structure and Evolution of Celestial Objects, Chinese Academy of Sciences, P.O. Box 110, Kunming 650216, People's Republic of China\\
            $^4$ Key Laboratory of Space Astronomy and Technology, National Astronomical Observatories, Chinese Academy of Sciences, Beijing 100101, People's Republic of China\\  
             $^5$ Center for Astronomical Mega-Science, Chinese Academy of Sciences, 20A Datun Road, Chaoyang District, Beijing, 100012, China\\    
             }

\abstract
   {Massive binary stars play a crucial role in many astrophysical fields. 
Investigating the statistical properties of massive binary stars is essential to trace the formation of massive stars and constrain the evolution of stellar populations. 
However, no consensus has been achieved on the statistical properties of massive binary stars, mainly due to the lack of a large and homogeneous sample of spectroscopic observations.}
   {We study the intrinsic binary fraction $f_{\rm b}^{\rm in}$ and distributions of mass ratio $f(q)$ and orbital period $f(P)$ of early-type stars (comprised of O-, B-, and A-type stars) and investigate their dependences on effective temperature $T_{\rm eff}$, stellar metallicity [M/H], and the projection velocity $v\sin{i}$, based on the homogeneous spectroscopic sample from the Large Sky Area Multi-Object Fiber Spectroscopic Telescope (LAMOST) Data Release Eight (DR8).}
    {We collected 886 early-type stars, each with more than six observations from the LAMOST DR8, and divided the sample into subgroups based on their derived effective temperature ($T_\mathrm{eff}$), metallicity ([M/H]), and projected rotational velocity ($v\sin{i}$). 
    Radial velocity measurements were archived from a prior study. A set of Monte Carlo simulations, following distributions of $f(P) \propto P^\pi$ and $f(q) \propto q^\gamma$ were applied to the observed binary fraction to correct for any observational biases.
    The uncertainties of the derived results induced by the sample size and observation frequency are examined systematically.
    }
    {We found that $f_{\rm b}^{\rm in}$ increases with increasing $T_\mathrm{eff}$. For stars in groups of B8-A, B4-B7, O-B3, the binary fractions are   
    $f_{\rm b}^{\rm in}=48\%\pm10\%$, $60\%\pm10\%$, and $76\%\pm10\%$, respectively. The binary fraction is positively correlated with metallicity for spectra in the sample, with derived values of $f_{\rm b}^{\rm in}=44\%\pm10\%$, $60\%\pm10\%$, and $72\%\pm10\%$ for spectra with metallicity ranges of [M/H]$<$-0.55, -0.55 $\leq$ [M/H]<-0.1, to [M/H] $\geq$ -0.1.
Over all the $v\sin{i}$ values we considered, the $f_{\rm b}^{\rm in}$ have constant values of $\sim$50\%.
It seems that the binary population is relatively evenly distributed over a wide range of $v\sin{i}$ values, while the whole sample shows that most of the stars are concentrated at low values of $v\sin{i}$ (probably from strong wind and magnetic braking of single massive stars) and at high values of $v\sin{i}$ (likely from the merging of binary stars). 
Stellar evolution and binary interaction may be partly responsible for this.
In the case of samples with more than six observations, we derived $\pi=-0.9\pm0.35$, $-0.9\pm0.35$, and $-0.9\pm0.35$, and $\gamma=-1.9\pm0.9$, $-1.1\pm0.9$, and $-2\pm0.9$ for stars of types O-B3, B4-B7, and B8-A, respectively.
    There are no correlations found between $\pi$($\gamma$) and $T_{\rm eff}$, nor for $\pi$($\gamma$) and [M/H].
The uncertainties of the distribution decrease toward a larger sample size with higher observational cadence.} 
    {}


\titlerunning{LAMOST OBA STARS} 
\authorrunning{Yanjun Guo et al.}        
\maketitle

\section{Introduction}\label{sec:Intro}
Many studies show that most stars are in binary systems \citep{1969Heintz,1976Abt,1991Duquennoy,2008Machida}, especially  early-type stars \citep{2012SanaScience,2012Chini,2013Duchene,2015Dunstall,2017MoeStefano}.
Statistical analyses of binary stars, 
including intrinsic binary fractions, distributions of orbital period, and  mass ratios, are important tracers of stellar formation and basic physical inputs for binary population synthesis \citep{2013sana,2019liuchaosmokinggun,2020Hanzhanwen}.
In particular, massive binaries and their statistical properties are essential for understanding compact objects' formation. 
These include double black holes, double neutron stars, and neutron star--black hole binaries, as they are the dominant gravitational wave sources for the  Laser Interferometer Gravitational-wave Observatory (LIGO), Virgo and KAGRA detectors \citep{2016AbbottGW,2016AbbottGW2,2018XueFei,2020Hanzhanwen,2020LangerGW}.

Among the statistical properties, the binary fraction $f_b$ is one of the most critical parameters, and it has been widely explored over the past few decades \citep{1983Carney,1990Henry,1991Duquennoy,1992Fischer,1998Mason,2002Latham,2010Kratter,2010Rastegaev,2010Raghavan,2013Duchene,2013sana,2014Tanaka,2017MoeStefano,2019liuchaosmokinggun}. 
These studies show that the binary fraction may depend on the mass, metallicity, and population age of the sample stars. 
The results from the various studies differ, and are controversial in some cases.
For example, \cite{2012SanaScience} reported that  Galactic O-type stars have an estimated binary fraction of 69\%, while \cite{2014Kobulnicky} suggested a binary fraction of 51\% for O-type stars using a different sample of stars. 
\cite{2002Latham} reported no correlation between the $f_b$ and metallicity for stars in their sample. 
Later works by \cite{2010Raghavan}, \cite{2018Tianzhijia}, and \cite{2019liuchaosmokinggun} suggest that an anti-correlation of the binary fraction and metallicity appears in their studies.
Alternatively, a positive correlation between these two quantities was reported from \cite{1983Carney} and \citet{2015Hettinger}. 

For the studies mentioned above, many works were based on small sample sizes, and on heterogeneous samples, especially for massive stars or early-type stars. 
The ongoing spectroscopic surveys, 
such as the Sloan Digital Sky Survey (SDSS) \citep{2000YorkSDSS} and the Large Sky Area Multi-Object Fiber Spectroscopic Telescope (LAMOST) \citep{2012CuiXiangQun,2020LiuChao}, 
provide good opportunities to study stellar statistical properties with huge homogeneous samples.  
\cite{2015Yuan} found an average binary fraction of $41\%\pm2\%$ for field FGK-type stars using spectroscopic sample from the SDSS.
The fractions (at least 10\%) of carbon-enhanced metal-poor (CEMP) stars and double-lined spectroscopic binaries, based on metal-poor stars selected from SDSS, are investigated by \cite{2015Aoki}.
Based on  low-resolution observation from the LAMOST DR5, \citet{2021luofeng} reported a binary fraction of 40\% for O- and B-type stars in the sample using spectra with more than three observations. Later work by \citet{2021GYJfb} utilizing the medium-resolution spectra with a cadence of more than two from the LAMOST DR7 suggests a binary fraction of 68\% for the early-type OB stars. 

In the work of \cite{2021GYJfb} (hereafter Paper I),
the relative radial velocity is obtained by a maximum likelihood estimation,
and a star is recognized as a binary 
according to the maximum variation of radial velocity. 
The intrinsic binary fraction can be obtained by correcting for any observational biases appearing in the sample using a series of Monte Carlo simulations from \citet{2013sana}. 
Since most of the early-type star spectra from \citet{2021GYJfb} only have two observations, significant errors would exist,  
although the large number of stars in the sample may reduce the errors 
to a certain degree.  
This influence has not been examined and corrected due to the small cadence number of spectra from LAMOST DR7. 
For the same reason, the dependence of the statistical properties on metallicity is not investigated in Paper I. 

Motivated by the recent release of more than six million medium-resolution spectra from LAMOST DR8, in this paper
we examine the uncertainties of the method 
used in Paper I and \cite{2013sana}, 
using both the sample size and the number of observations. 
We aim to improve the investigation of the binary fraction for early-type stars in the LAMOST DR8 database using an increased sample size (886 stars) with a higher observational cadence ($\ge 6$) to determine the dependences of the statistical properties of these stars on their metallicity.

The structure of the paper is as follows. 
We introduce the data of LAMOST-MRS in Section \ref{sec:LAMOST DATA}. 
In Sect.~\ref{sec:Sample selection and RV measurements} we describe our work of dividing the sample into different groups based on their parameters and radial velocity (RV) measurements, and the criterion for identifying binaries.
We briefly describe the Monte Carlo method used to correct the observational bias, the consistency check, and the details of testing the suitability of this method in Section~\ref{sec: Correction for observational biases}. 
We report the results of the relationships between the statistical properties and the stellar spectral types, as well as the metallicity for early-type belonging to various subgroups from LAMOST DR8 in Sect.~\ref{sec: Results and Discussion}. 
The summary and our conclusions are presented in Sect.~\ref{sec:Conclusion}.

\section{LAMOST data and sample}\label{sec:LAMOST DATA}
LAMOST is a four-meter quasi-meridian reflecting Schmidt telescope located at the Xinlong station of the National Astronomical Observatory, implemented with 4000 fibers. 
Both medium-resolution spectrographs with a resolving power of $R\sim 7500$ and low-resolution spectrographs with a resolving power of $R\sim 1800$ were installed on the telescope \citep{2012CuiXiangQun,2012ZhaoGang,2012DengLiCai}. 
In October 2018, LAMOST began a new medium-resolution survey (MRS) to obtain multi-epoch observations. The spectra of MRS observations made from the blue arm have a wavelength range of $495\sim535$~nm, and cover a wavelength range of $630\sim680$~nm for the red arm \citep{2020LiuChao}. 

\cite{2021GYJslam} utilized the spectra from the LAMOST MRS database to derive the atmospheric parameters of 9,382 early-type stars via a data-driven technique called stellar label machine (SLAM) \citep{2020zhangbolaspecslam,2020zhangboMRS}. 
In this study, we adopt the early-type stars from \cite{2021GYJslam} to investigate their statistical properties.

\section{Sample selection and RV measurements}\label{sec:Sample selection and RV measurements}
\subsection{Classifying the sample}\label{sec: Classify the sample}

\begin{table}
\caption{\label{tab:grouped sample size}Sample size of each group.}
\centering
\begin{tabular}{lc}
 \hline \hline
 Group  & Number  \\  
 \hline                   
 B8-A &592\\
 \hline
 B4-B7 & 140\\
 \hline
 O-B3 &154\\
 \hline
Metal-poor: [M/H]<-0.55 & 612\\
\hline
Metal-medium: -0.55 $\leq$ [M/H]<-0.1 & 143\\
\hline
Metal-rich: [M/H] $\geq$ -0.1    & 131\\
\hline
Low $v\sin{i}$: $v\sin{i}$ $<$ 35 $\ {\rm km\ s^{-1}}$  &595 \\
\hline
Medium $v\sin{i}$: 35 $\leq$ $v\sin{i}$<70    &176\\
\hline
High $v\sin{i}$: $v\sin{i}$ $\ge $ 70 $\ {\rm km\ s^{-1}}$ &115 \\ 
\hline  \hline
\end{tabular}
\end{table}

As mentioned in Sect.~\ref{sec:Intro}, both the sample size and observation frequency affect the uncertainty of binary fraction and the distributions for mass ratio and orbital periods.
In the work of \cite{2013sana} about 93\% of the samples have more than six observations. 
\cite{2021luofeng} found that an increase in the observational frequency of a sample may lead to smaller uncertainties.
We thus selected the sample with more than six observations
from the catalog of \cite{2021GYJslam} for our study. 
There are a total of 886 stars in the sample.
We also did some tests on the effect of observational frequencies and sample size to examine the uncertainties of the method, and the results are shown in Sect.~\ref{sec: Results and Discussion}.

\begin{figure}
       \centering
       \includegraphics[scale=0.4]{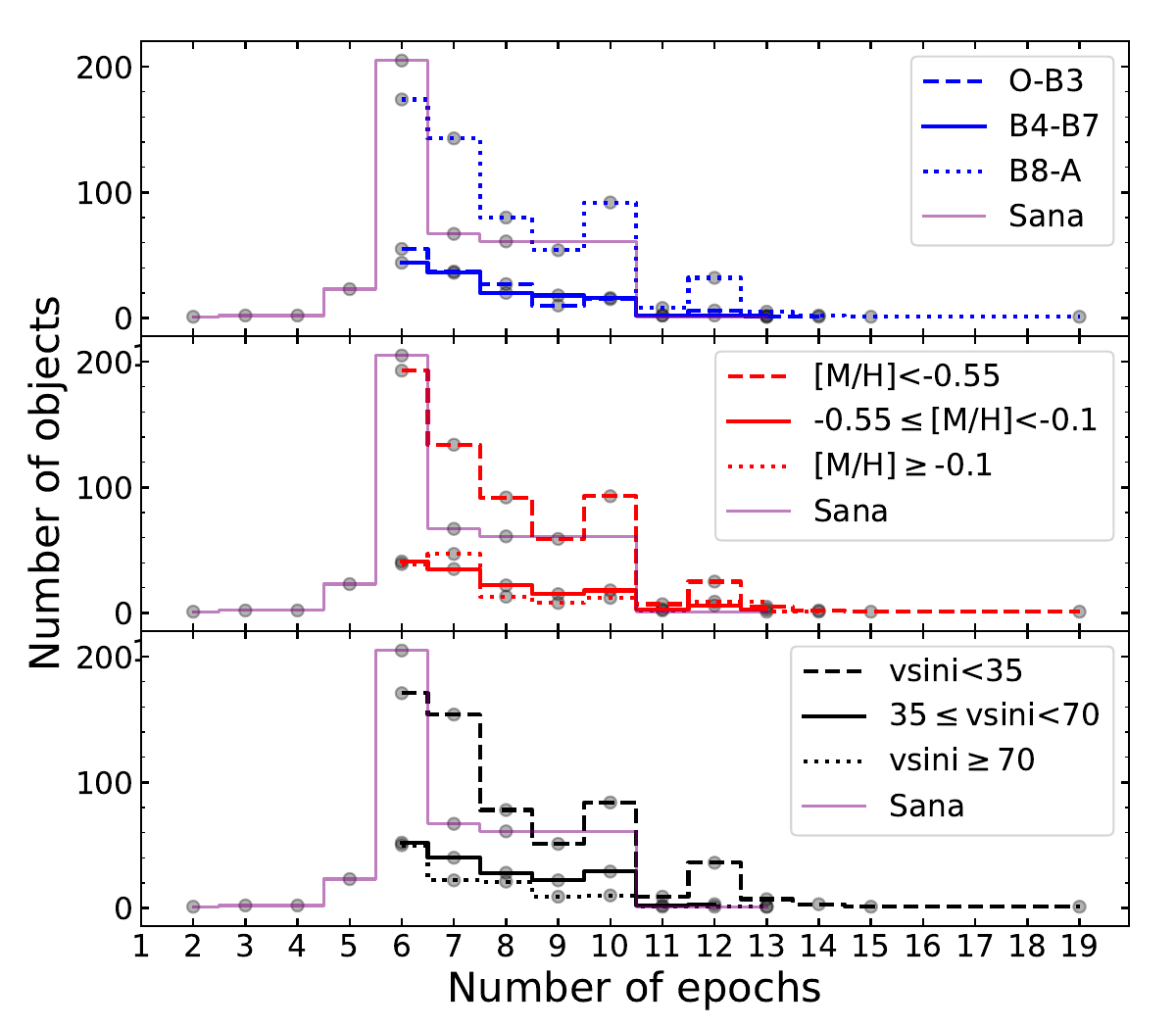}
        \caption{Number distribution for different groups based on $T_\mathrm{eff}$ (top panel), [M/H] (middle panel), and $v\sin{i}$ (bottom panel). The solid purple line represents the observations of samples from \cite{2013sana} in which about 93\% of the sample are from more than six observations.}\label{fig: obs time}
\end{figure}

Based on the atmospheric parameters given by \cite{2021GYJslam}, we 
divided the samples into different groups according to three observables: effective temperature ($T_\mathrm{eff}$), metallicity ([M/H]), and projected rotational velocity ($v\sin{i}$). 
We simply divided the sample stars into low, medium, and high groups based upon their spectral types, metallicity, and $v\sin{i}$ because the aim of the grouping is to explore trends between those atmospheric parameters and the binarity of the samples. 
For $T_\mathrm{eff}$, we divided the 886 samples into three groups (B8-A, B4-B7, O-B3)\footnote{We followed the same boundaries as used in \cite{2021GYJfb} to facilitate comparison with the previous results. 
There are four groups according to equivalent widths of He I and H$\alpha$: T1 ($\sim$ O-B4), T2 ($\sim$ B5), T3 ($\sim$B7), and T4 ($\sim$B8-A). We had not obtained effective temperature at that time. 
However, we found that the number in the first three groups (T1-T3, with six observations) 
is not large enough for the study. 
We therefore combined the first three groups and divided them into two groups, 
with B3-B4 as the boundary to ensure the two groups have similar samples.}
to investigate the relationships between $T_\mathrm{eff}$ and statistical properties. 
We display the number distribution of stars classified by spectral types of B8-A (dotted blue line), B4-B7 (solid blue line), and O-B3 (dashed blue line) in the top panel of Figure~\ref{fig: obs time}. 
The sample distribution from \citet{2013sana} is denoted in solid purple. 
For [M/H], we divided the 886 stars into groups of [M/H]<-0.55 (red dashed line), -0.55 $\leq$ [M/H]<-0.1 (solid red line), [M/H] $\geq$ -0.1 (red dotted line), shown in the middle panel of Fig.~\ref{fig: obs time}.
The black dashed line and the black dotted line represent the stars with $v\sin{i}$ $<$ 35 $\ {\rm km\ s^{-1}}$ and the stars with $v\sin{i}$ larger than 70 $\ {\rm km\ s^{-1}}$, respectively, in the bottom panel of Fig.~\ref{fig: obs time}.
The number distribution of stars with 35 $\leq$ $v\sin{i}$<70$\ {\rm km\ s^{-1}}$ is shown in the solid black line.
The grouped sample sizes are displayed in Table~\ref{tab:grouped sample size}.

\subsection{Radial velocity measurements} 
\begin{table*}
\caption{\label{tab:RV Catalogs} Radial velocity catalogs.}
\centering
\begin{tabular}{lccccccccccl}
 \hline \hline
Star  & RA    & DEC    &  Date   & S/N & $T_\mathrm{eff}$  & $[M/H]$  &  $v\sin{i}$    &     $RV$      & $\sigma$\\OBSID & (deg) & (deg)  & (MJD)   &     &      (K)          &  (dex)   & ($\ {\rm km\ s^{-1}}$)  & ($\ {\rm km\ s^{-1}}$) & $\ {\rm km\ s^{-1}}$)\\ \hline
693504025    &33.1157  &58.6092  & 58446.8958   &157  &21579 &-0.25  &41   &-96.90    &0.64\\
715104025    &33.1157  &58.6092  & 58494.8132   &135  &21579 &-0.25  &41   &-96.42    &0.39\\
685011073    &35.0524  &60.0624  & 58421.0583   &168  &22265 &-0.28  &53   &-56.78    &0.42\\
702711073    &35.0524  &60.0624  & 58468.825    &163  &22265 &-0.28  &53   &-55.20    &0.42\\
702712188    &36.9131  &60.2170  & 58468.841    &115  &18023 &-0.46  &135  &-56.41    &1.22\\
685012188    &36.9131  &60.2170  & 58421.0417   &115  &18023 &-0.46  &135  &-52.87    &1.25\\
\hline  \hline
\end{tabular}

\tablefoot{This table is available in its entirety in machine-readable form. The typical uncertainties of $T_\mathrm{eff}$, [M/H], and $v\sin{i}$ are 
2,185K, 0.2 dex,\footnote{Here we use the SLAM (model) error at the S/N=20 (the maximum value of error), as shown in Fig. 4 of \cite{2021GYJslam}. The high-resolution comparison spectra give individual element abundances rather than [M/H]. \cite{2021GYJslam} therefore did not give a typical uncertainty of [M/H].} and 11$\ {\rm km\ s^{-1}}$, respectively.}
\end{table*}

The data reduction was carried out via the standard LAMOST 2D pipeline, and the wavelength calibration was accomplished using the Sc and Th-Ar lamps for MRS spectra. 
The typical accuracy of wavelength calibration is 0.005nm\ pixel$^{-1}$ for LAMOST-MRS \citep{2015luoaliLRSstellarpara,2021GYJfb,2021Renjuanjuanwvcalu}. 

However, \cite{2019LiunianRV} and \cite{2021zhangboRV} concluded that significant systematic errors exist among the RVs obtained from spectra collected by the different spectrographs, the exposures of LAMOST MRS surveys, and the temporal variations of the RV zero-points. 
Therefore, \cite{2021zhangboRV} applied a robust self-consistent method using \textit{Gaia} DR2 RVs to determine the RV zero-points (RVZPs) from the exposures for each spectrograph of LAMOST after measuring the RVs based on the cross-correlation function method. 

We cross-matched our 886 early-type stars with the updated RV catalog\footnote{\url{https://github.com/hypergravity/paperdata}} for LAMOST DR8 from \cite{2021zhangboRV} and adopted their corrected RV measurements. In Table~\ref{tab:RV Catalogs}, for each star we list the observation ID,   coordinates, MJD date, S/N, $T_\mathrm{eff}$, [M/H], $v\sin{i}$, RV, and the uncertainty ($\sigma_{i}$).

\subsection{Criterion for the binary} 
To identify the binaries in our sample, we adopted Equation 4 from \cite{2013sana},
which states that a star is a binary if its RVs satisfy

\begin{equation}
\centerline{ $\frac{|v_{i} - v_{j}|}{\sqrt{\sigma_{i}^2 \ + \sigma_{j}^2}}\  > \ 4$ and ${|\\ v_{i} - v_{j} \\ |}\  >  \ C$,}\label{cer:SB1}
\end{equation}
where $v_{i(j)}$ is the RV measured from the spectrum at epoch i~(j) and $\sigma_{i(j)}$ is the associated uncertainty. This criterion has been applied in several similar studies, for example \cite{2012SanaScience}, \cite{2015Dunstall}, \cite{2021luofeng},\cite{2021MahyC}, \cite{2021Banyard}, and \cite{2021GYJfb}. 

According to statistical standards, the confidence threshold of 3$\sigma$ suggests a probability of finding 30 false detections in a sample of 1000 data entries, and this threshold is sufficient to filter out any outliers. 
We adopted a  stricter criterion of 4$\sigma$, as suggested by \citet{2013sana}. 
This threshold criterion indicates a probability of finding one false detection in every 1000 entries.

The threshold C is used here to filter stars with pulsations, 
which may also lead to significant RV variation. 
\cite{2013sana} adopted C=20~$\ {\rm km\ s^{-1}}$ for O-type stars, 
and \cite{2015Dunstall} adopted C=16 $\ {\rm km\ s^{-1}}$ for B-type stars, 
based on the kinks in the RV distributions of their sample.
We did not find   the appearance of a kink feature in our sample, 
hence simply adopted C= 16 $\ {\rm km\ s^{-1}}$ as that in \cite{2015Dunstall}
because our sample is dominated by B-type stars.

\begin{figure}
        \centering
            \includegraphics[scale=0.6]{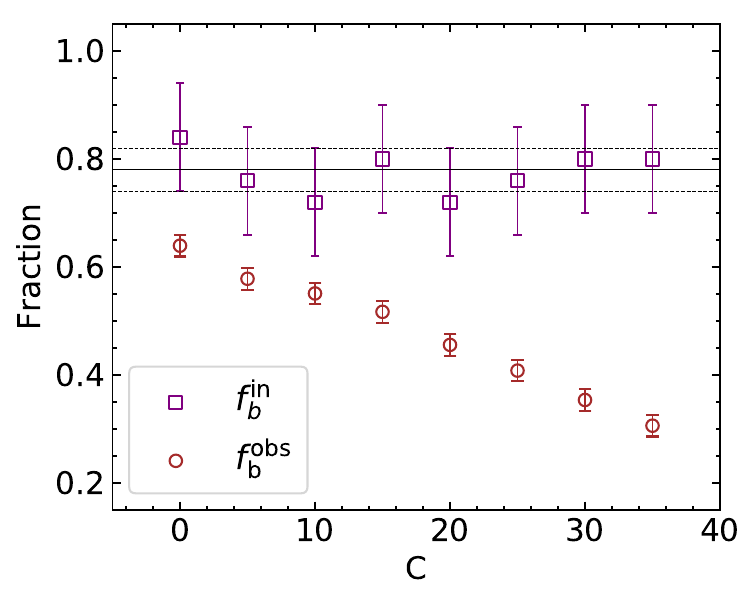}
                \caption{Variations in the binary fraction of observation ($f_{\rm b}^{\rm obs}$) and simulation ($f_{\rm b}^{\rm in}$) filtering from a set of different $C$ values. The red circles and squares represent the binary fraction of the observation and the intrinsic binary fraction after Monte Carlo simulation correction, respectively. The gray dashed and solid lines represent the mean and standard deviation of these $f_{\rm b}^{\rm in}$, respectively.}\label{fig: testC}
\end{figure}

The observed binary fraction $f_{\rm b}^{\rm obs}$ could be underestimated by the constraint of C because of the binary stars, even with large RV amplitudes, but small phase differences of the spectroscopic observations might be missed, and similarly binaries with relatively small RV variations might also be missed.
Moreover, a large value of C would lead to a small value of $f_{\rm b}^{\rm obs}$ from the same sample. 
However, this biases could be corrected by the Monte Carlo simulation described in Sect.~\ref{sec: Correction for observational biases}. 
Here we just show (see Fig.~\ref{fig: testC}) the results of the corrections for the O-B3 group. 
In the figure, the circles represent the observed binary fractions ($f_{\rm b}^{\rm obs}$) of binaries in the sample (i.e.,\ without applying any corrections), and these distributions are heavily affected by the arbitrary choice of $C$ values. 
The squares stand for the intrinsic binary fractions ($f_{\rm b}^{\rm in}$) after correcting any observational biases using a Monte Carlo simulation (discussed in Sect.~\ref{sec: Monte-Carlo method}). 
As shown in the figure, the intrinsic binary fractions are constantly distributed over the $C$ values, suggesting that it is safe to adopt the $C=16$ km s$^{-1}$ threshold criterion \citet{2021MahyC}.

\section{Correction for observational biases}\label{sec: Correction for observational biases} 
\subsection{Monte Carlo method}\label{sec: Monte-Carlo method} 
Here we use the approach described in \cite{2013sana},
where several Monte Carlo (MC) simulations are run to assess the intrinsic binary fraction ($f_{\rm b}^{\rm in}$) based on $f_{\rm b}^{\rm obs}$.
To perform the simulations, we need to construct two synthetic cumulative distributions (CDF) of RV variation ($\Delta$RV), the peak-to-peak RV variation of each star, and the minimum timescale between the exposures ($\Delta$MJD) by assuming the orbital configurations.
We use a power law to describe the distribution for both the orbital period $P$ and mass ratio $q$: $f(P) \propto P^\pi$ and $f(q) \propto q^\gamma$. 
We do not use (\rm log $P$) $\propto$ (log\ $P)^\pi$, as is done in the literature, because the linear form is more sensitive to short-period binaries \citep{2021GYJfb}.
Table~\ref{tab:sim parameter} lists the variable ranges for the above parameters and power index. 
We adopt the initial mass function (IMF) from \cite{1955Salpeter} with $\alpha$ $=$~$-$2.35\footnote{We did a test with the IMF of \cite{2001Kroupa} (i.e., alpha = -2.3 for M$\ge$1Msun) 
and obtained binary fractions similar to those shown in the paper.} for the primary masses \citep{2021MahyC}.

In general, two-body kinetic systems in a Keplerian orbit are described by the orbital parameters, which are inclination (i), semimajor axis (a), the argument of periastron ($\omega$), eccentricity (e), and the epoch of periastron ($\tau$). 
The inclination is randomly drawn over an interval from 0 to $\pi$/2 and satisfies a probability distribution of $sin(i)$. 
The semimajor axis is correlated to the orbital period (P). 
The $\omega$ satisfies a uniform distribution randomly drawn from it from 0 to 2$\pi$. 
We use $e^\eta$ to describe the distribution in which $\eta$ is set to -0.5 from \cite{2013sana}, and $\tau$ is selected randomly in units of days.

\begin{figure}
        \centering
            \includegraphics[scale=0.8]{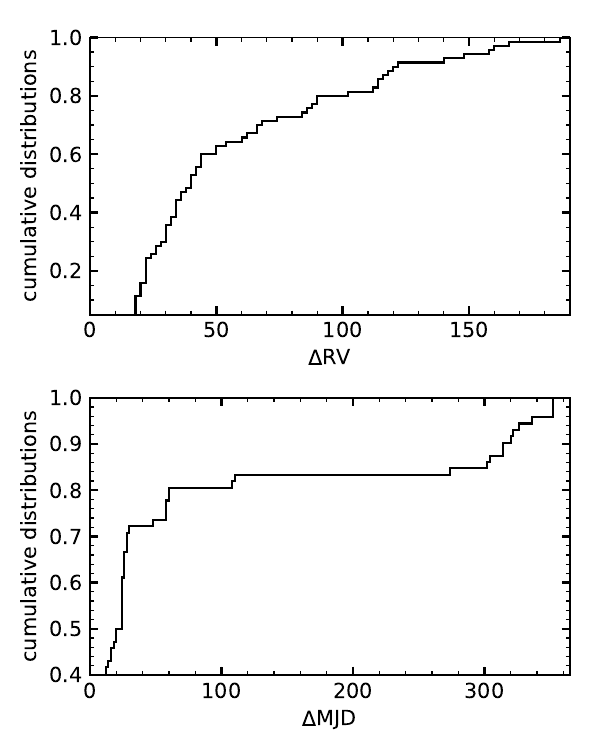}
                \caption{Example of the simulated CDF of $\Delta$RV and $\Delta$MJD with the input parameters of $f_{\rm b}^{\rm in}$=0.72, $\pi$=-0.9, and $\gamma$=-1.9.}\label{fig: exampleCDF}
\end{figure}

With the above assumptions we can obtain the simulated RV to further simulate the CDF distribution of $\Delta$RV and $\Delta$MJD. The example for the simulated CDF of $\Delta$RV and $\Delta$MJD is shown in Fig.~\ref{fig: exampleCDF}.
The constructed simulated CDFs are compared with the observed ones by using Kolmogorov--Smirnov (KS) test, and the simulated and observed
fractions of detected binaries ($f_{\rm b}^{\rm sim}$ and $f_{\rm b}^{\rm obs}$) are compared using binomial distribution \citep{2013sana}.
We then use the optimum values from the global merit function (GMF) projection, following \cite{2013sana}, to estimate the final results.
The GMF consists of the KS probabilities of $\Delta$RV and $\Delta$MJD distributions and the binomial probability 
\begin{equation}
\centerline{ GMF= $P_{ks}$($\Delta$RV) $\times $ $P_{ks}$($\Delta$MJD) $\times $ $B(N_{b},N,f_{\rm b}^{\rm sim}$),}
\end{equation}
where $N_{b}$ is the number of binaries in the observations, while $N$ is the sample size.

\begin{figure}
        \centering
            \includegraphics[scale=0.4]{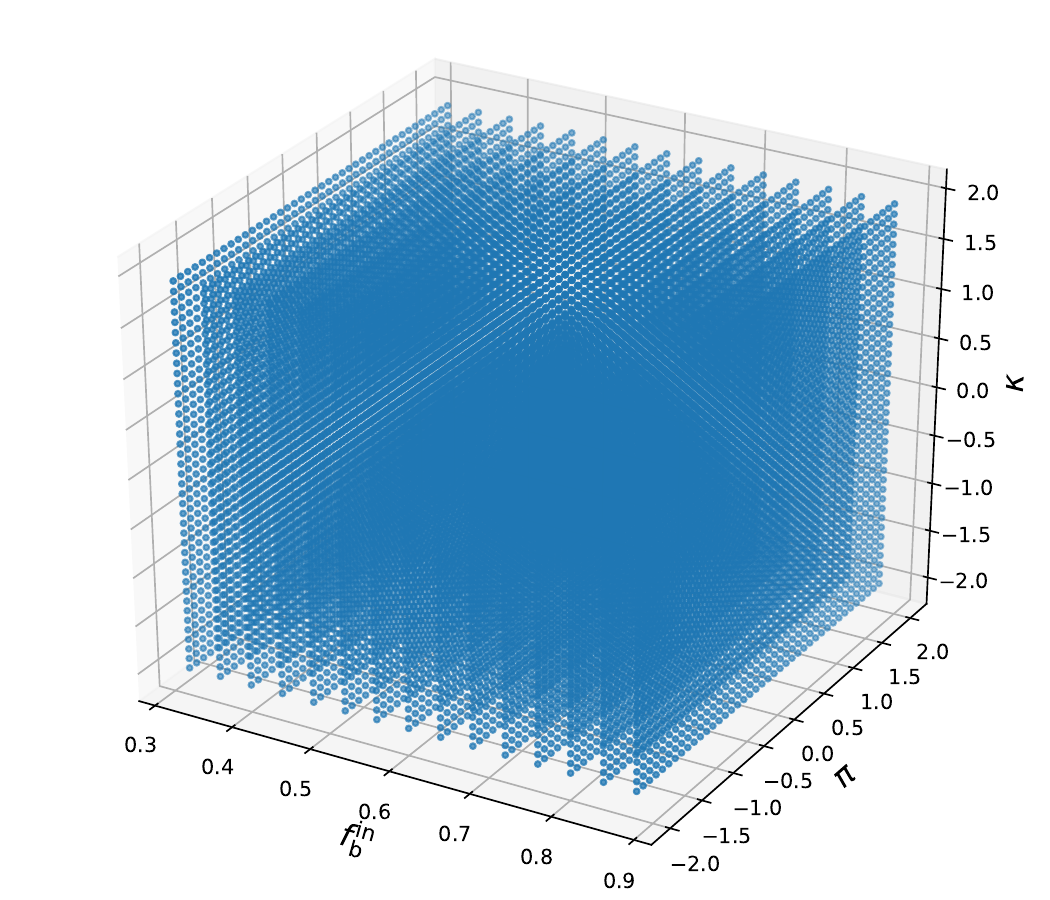}
                \caption{Choices of input sets of $f_{\rm b}^{\rm in}$, $\pi$, and $\gamma$ covering the range $f_{\rm b}^{\rm in}$ from 0.32 to 0.88 K with steps of 0.04 K, and $\pi$ and $\gamma$ from -2 to 2 with steps of 0.1.}\label{fig: simparameter}
\end{figure}

\begin{figure}
        \centering
            \includegraphics[scale=0.4]{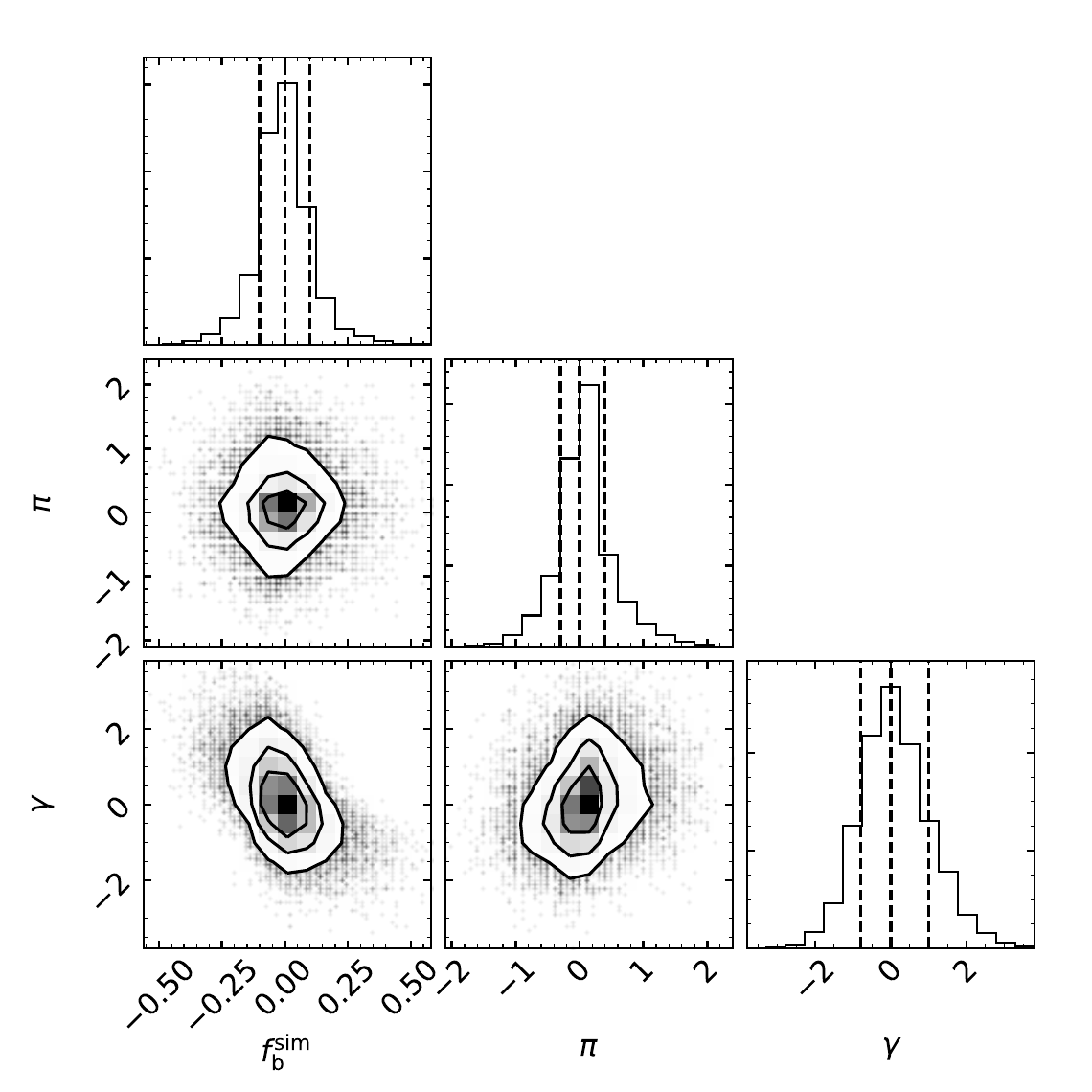}
                \caption{Projections of residuals from the comparisons of input known parameters and output parameters.}\label{fig: corner}
\end{figure}

To examine the applicability of the GMF for stars (more than six observations) from the LAMOST-MRS, we constructed 24600 synthetic CDFs with the known input sets of $f_{\rm b}^{\rm in}$, $\pi$, and $\gamma$.
Figure~\ref{fig: simparameter} shows the choices of the known input sets of the parameters.
We then compare the input parameters with the simulated parameters and show
the projections of residuals for the comparisons in Fig.~\ref{fig: corner}.
The residuals are the differences between the output parameters predicted by the simulations and the input known parameters. 
The projection is the distribution of the residuals in which the histogram is the marginal distribution for each parameter.
For each parameter, we use the 50th percentile of marginal distribution from the residual distributions to estimate offset and mean values of the 16th percentile (p16) and the 84th percentile (p84) to estimate errors. 
We estimated the uncertainty (unc) using the offset and error of each parameter through error propagation: unc=$\sqrt{(\frac{\rm p84 - \rm p16}{2})^2+ \rm p50^2}$.
The final uncertainties are $f_{\rm b}^{\rm in}$=0.1, $\pi$=0.35, and $\gamma$=0.9. 
We see in the figure that the intrinsic properties (input) are well reproduced by the Monte Carlo simulation without systematic bias (p50=0), 
which verifies the reliability of our method.  

We also reanalyzed the 360 O-type stars in the sample of \cite{2013sana}. We obtained statistical properties of the sample consistent with those of \cite{2013sana} (see Paper I for details).

\begin{table*}
\caption{\label{tab:sim parameter}Range of different parameters and power indexes used in MC simulation. The column Power Index shows 
$\pi$ and $\gamma$, the power index of $P$ (orbital period) and $q$ (mass ratio), respectively.
$f_{\rm b}^{\rm in}$ is the binary fraction for simulation.
The ranges of $P$ and $q$ are shown in the column Parameter Range, while the ranges of the power index are shown in the column Index Range.
The last row gives the step of each power index.}
\centering
\begin{tabular}{lccccl}
 \hline \hline
Parameter  & Power law  &  Parameter Range  &  Power Index &  Index Range &Step\\\hline
$P$(d)   &$f(P) \propto P^\pi$    & 1 - 1000        &$\pi$    &-2.50- 2.50      &0.1\\
$q$      &$f(q) \propto q^\gamma$ & 0.1 - 1.0       &$\gamma$ &-4   - $1.00^*$  &0.1\\
$f_{\rm b}^{\rm in}$  &-                       &-                &-        &0.20 - 1.00      &0.04\\
  \hline  \hline
\end{tabular}

\tablefoot{As for stars in group O-B3, the range of $\gamma$ is -2.5 - 2.50 since  earlier types are assumed to have slightly larger values \citep{2013sana,2015Dunstall}.}
\end{table*}

\subsection{Impact of observational frequency and sample size}\label{sec:test of observational frequencies} 
In order to investigate the effect of observational frequencies and the sample sizes on the final results, we performed the three sets of tests. 

\begin{figure}
        \centering
            \includegraphics[scale=0.4]{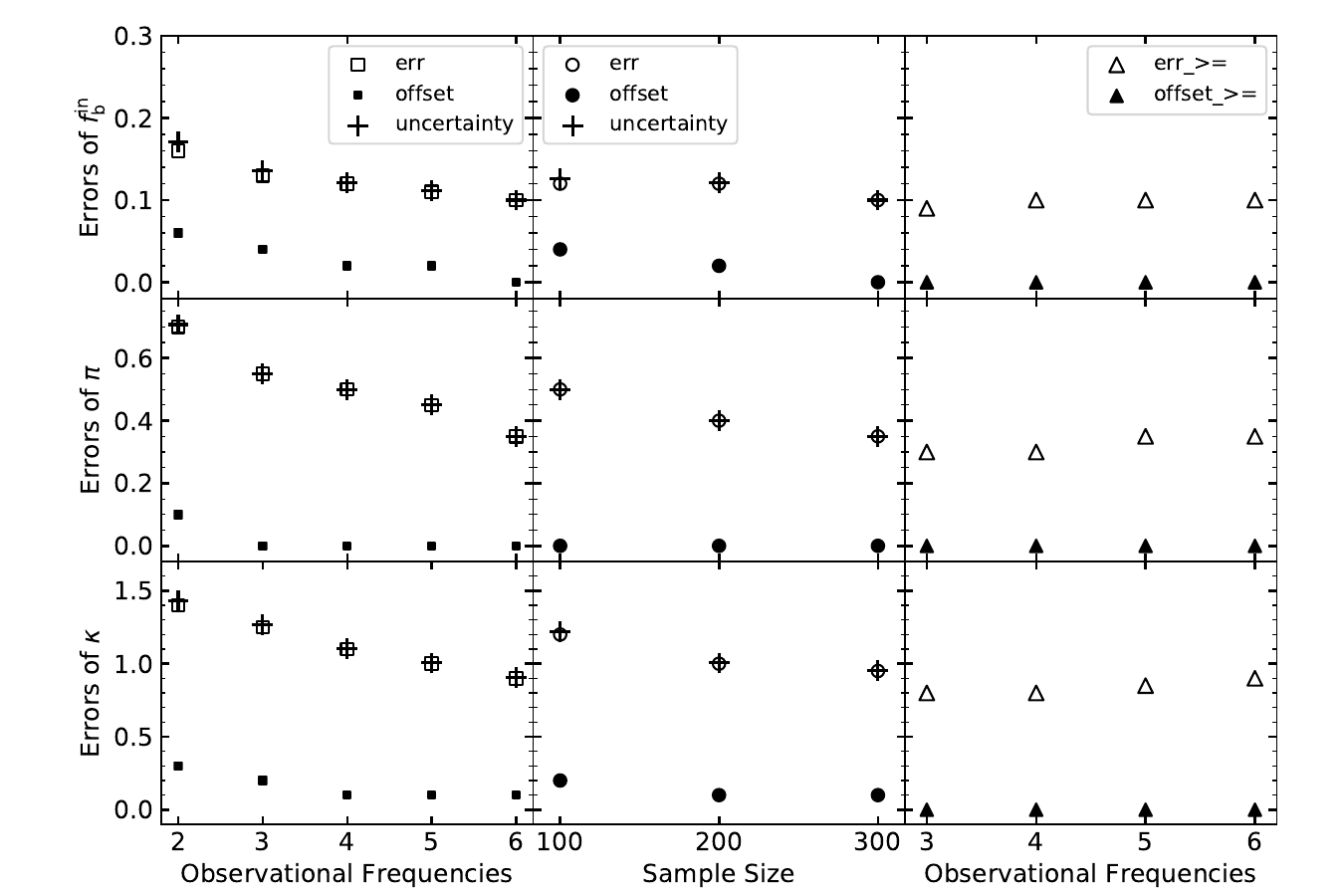}
                \caption{Uncertainties for the different samples. 
                Left panel: Open and filled squares represent the error and offset of the same sample of stars, but with different observational frequencies. The plus signs represent uncertainties.
Middle panel: Open and filled circles represent the error and offset of the same observational frequencies sample of stars, but with different sample sizes. The plus signs represent uncertainties.
                Right panel: Open and filled triangles represent the error and offset of the actual samples. The uncertainties are not shown because they have the same values as the errors.}\label{fig: errall}
\end{figure}

First, we examined the impact of observation frequency 
by using the same sample of stars ($N=154$, the same  sample size as O-B3), but with different observational frequencies $n$ (i.e., $n=2, 3, 4, 5$ and 6).
The result is shown in the left panel of Fig.~\ref{fig: errall}, where the black open and filled squares (plus signs) represent the error and offset (uncertainties). More detailed calculations are described in Sect.~\ref{sec: Monte-Carlo method}.
We see that the uncertainty of each parameter becomes smaller with 
increasing observational frequency, as expected \citep{2021luofeng}. 
In particular, the offset for all three parameters becomes zero when $n=6$.

We then investigated the effect of the sample size. 
We used three samples with the number of stars $N$=100, 200, and 300, and the observational frequency $n$ set to 5 (the result of $n$=6 is similar to $n$=5; here we take 5 as an example). 
The middle panel of Fig.~\ref{fig: errall} shows the results. 
It indicates that the uncertainty decreases with the increasing number of stars, especially for $\pi$ and $\gamma$.

We find that both big sample size and a high observational frequency would reduce the uncertainty of the method. However, for the real data 
the sample size generally decreases with increasing observational frequency.
We therefore make the uncertainty estimates using the number of observations and the sample size of the real sample. 
We consider four samples here,  with $n\ge$ 3, 4, 5, and 6. 
The number of stars in each sample is N=459, 300, 240, and 154.  
The results are shown in right panel of Fig.~\ref{fig: errall}.
We see that the uncertainties of binary fraction $f_{\rm b}^{\rm in}$ are very similar
(around 0.1) for all the cases,
while the uncertainties of $\pi$ and $\gamma$ are slightly smaller for the cases of $n\ge 3, 4$ than that for the cases of $n\ge 5, 6$.
The sample size could be the cause for this. 
This test indicates that the number of stars in the sample may indeed reduce the uncertainty of the results induced by the observational frequency.
In the following, we show all the results from the samples with observational frequencies greater than 3, 4, 5, and 6, 
and selected the case of $n\ge 6$ as the standard to compare with previous studies.

\section{Results and discussions}\label{sec: Results and Discussion}

Figures~\ref{fig: obsallresult_T} to \ref{fig: resultvsini} show the intrinsic statistical properties of $f_{\rm b}^{\rm in}$, $\pi$, and $\gamma$ for various samples, with observation frequency $n\ge$ 3, 4, 5, and 6, and the dependences of these parameters on the spectral type, metallicity, and projection velocity $v\sin{i}$.  

\subsection{Dependence on $T_{\rm eff}$}\label{sec: teff}
\begin{figure*}
        \centering
            \includegraphics[scale=0.6]{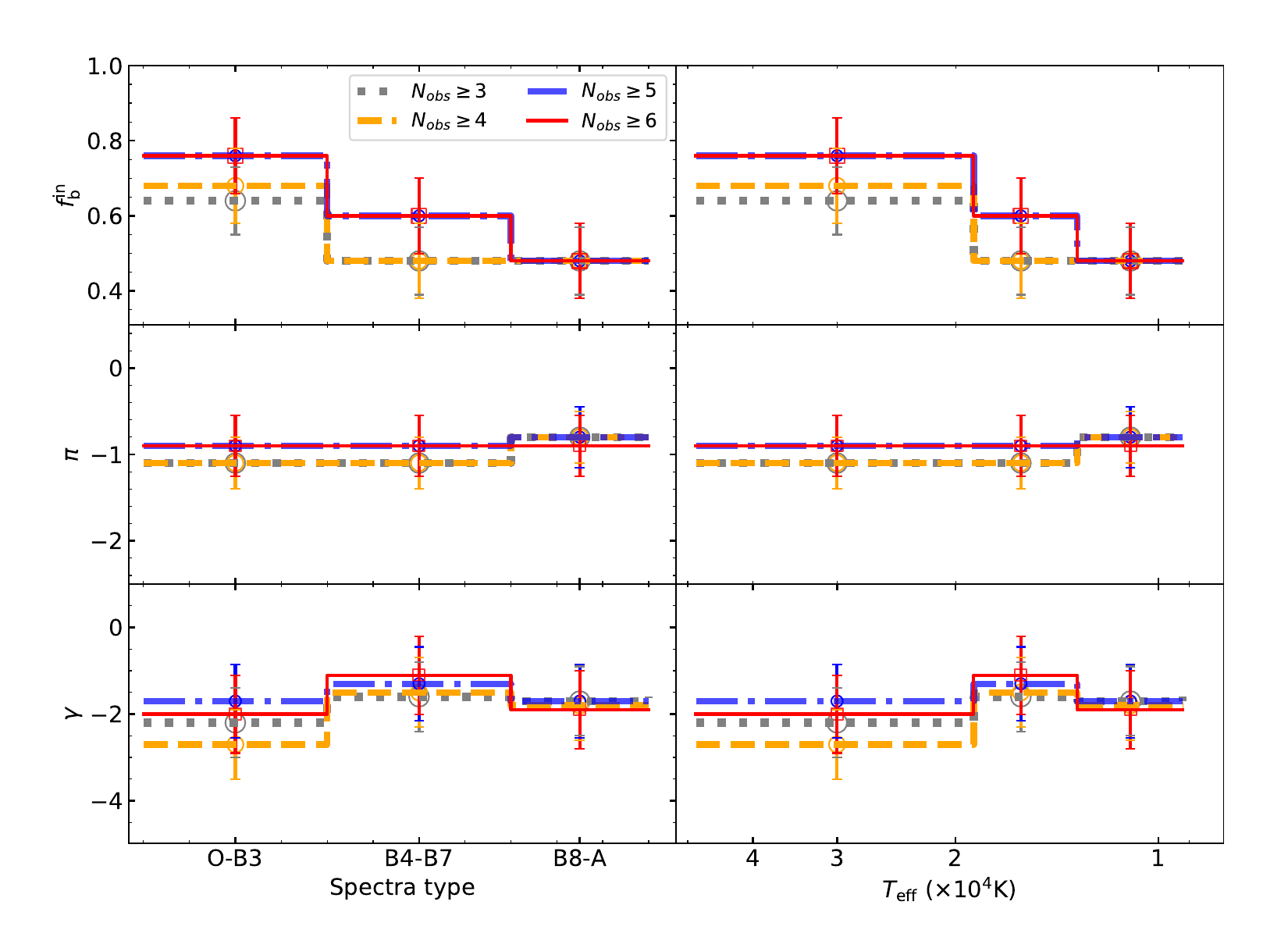}
                \caption{Intrinsic statistical properties of $f_{\rm b}^{\rm in}$, $\pi$, and $\gamma$ for stars with observational frequencies greater than 3, 4, 5, and 6 in groups of classification based on $T_{\rm eff}$. The x-axis in the left panel is based on spectra type, while in the right panel it is based on $T_{\rm eff}$. The top panel represents the trend of $f_{\rm b}^{\rm in}$, and the middle (bottom) represents the trend of $\pi$ ($\gamma$). The blue dot-dashed and red solid lines represent the results of stars classified based upon $T_{\rm eff}$ in groups of B8-A, B4-B7, and O-B3  with observational frequencies greater than five and six, respectively, while the gray and yellow dashed lines represent frequencies greater than  three and four.}\label{fig: obsallresult_T}
\end{figure*}

Figure~\ref{fig: obsallresult_T} shows the dependence of $f_{\rm b}^{\rm in}$, $\pi$, and $\gamma$ on the spectral type.
We note that the intrinsic binary fraction $f_{\rm b}^{\rm in}$ is positively correlated with $T_{\rm eff}$ (i.e., the intrinsic binary fractions increased toward the early-type stars), which is consistent with the result of \cite{2010Raghavan}, \cite{2013Duchene}, \cite{2017MoeStefano}, and \cite{2021GYJfb}. 
\cite{2011MarksandKroup} also argued that binaries born preferentially efficiently with larger primary mass \citep{2019liuchaosmokinggun}. 

For the cases of $n \ge 5$ and 6, the binary fraction is $76\%\pm10\%$, $60\%\pm10\%$, and $48\%\pm10\%$ for stars of O-B3, B4-B7, and B8-A types, respectively. 
The binary fraction for stars of types B4-B7 and B8-A in the cases of $n \ge 3$ and 4 becomes smaller ($f_{\rm b}=48\%\pm10\%$) which is expected.
For a large observation frequency (i.e., $n\ge5$ or 6), there is a high possibility of obtaining the maximum RV difference of a binary, resulting in more binaries to be verified in comparison to the cases of relatively low observation frequencies. The results for $n\ge5$ and 6 are very similar, indicating that 5 and 6 times are the optimal numbers of the observational frequencies for this method.

Figure~\ref{fig: result_T} shows the $f_{\rm b}^{\rm in}$ for the sample of $n\ge6$ in comparison to that in the literature.
We see that the result of O-B3-type stars in our study agrees well with that  
of Galactic O-type stars ($f_{\rm b}^{\rm in}=69\% \pm 9\%$, filled squares) from \cite{2012SanaScience}, 
and that $f_{\rm b}^{\rm in}$ in our sample is significantly higher than that given by \cite{2021luofeng} for the OB stars from LAMOST DR5 (about 40\%, with filled circles). 
In the study of \cite{2021luofeng}, most of stars (80\%) in the sample have three observations. 
As seen in Fig.~\ref{fig: obsallresult_T}, the binary fraction is about 48\% for B4-A stars for the sample of $n\ge 3$, 
which is consistent with the result of \cite{2021luofeng} within the uncertainties.
The binary fraction in \cite{2021GYJfb} is systematically lower than that of this study for each spectral type 
since most of the samples in \cite{2021GYJfb} have only two observations, 
leading to many binaries being missed by the method, as mentioned in Sect.~\ref{sec: Correction for observational biases}.

The values of $\pi$ and $\gamma$ remain almost constant with the spectral type of stars, 
possibly indicating the same origins of the binary star formation with various temperatures in this range. 
Our study gives $\pi=-0.9\pm0.35$, $-0.9\pm0.35$, and $-0.9\pm0.35$, and $\gamma=-1.9\pm0.9$, $-1.1\pm0.9$, and $-2\pm0.9$ for stars of O-B3, B4-B7, and B8-A type, respectively.

\subsection{Dependence on [M/H]}
\begin{figure}
        \centering
            \includegraphics[scale=0.6]{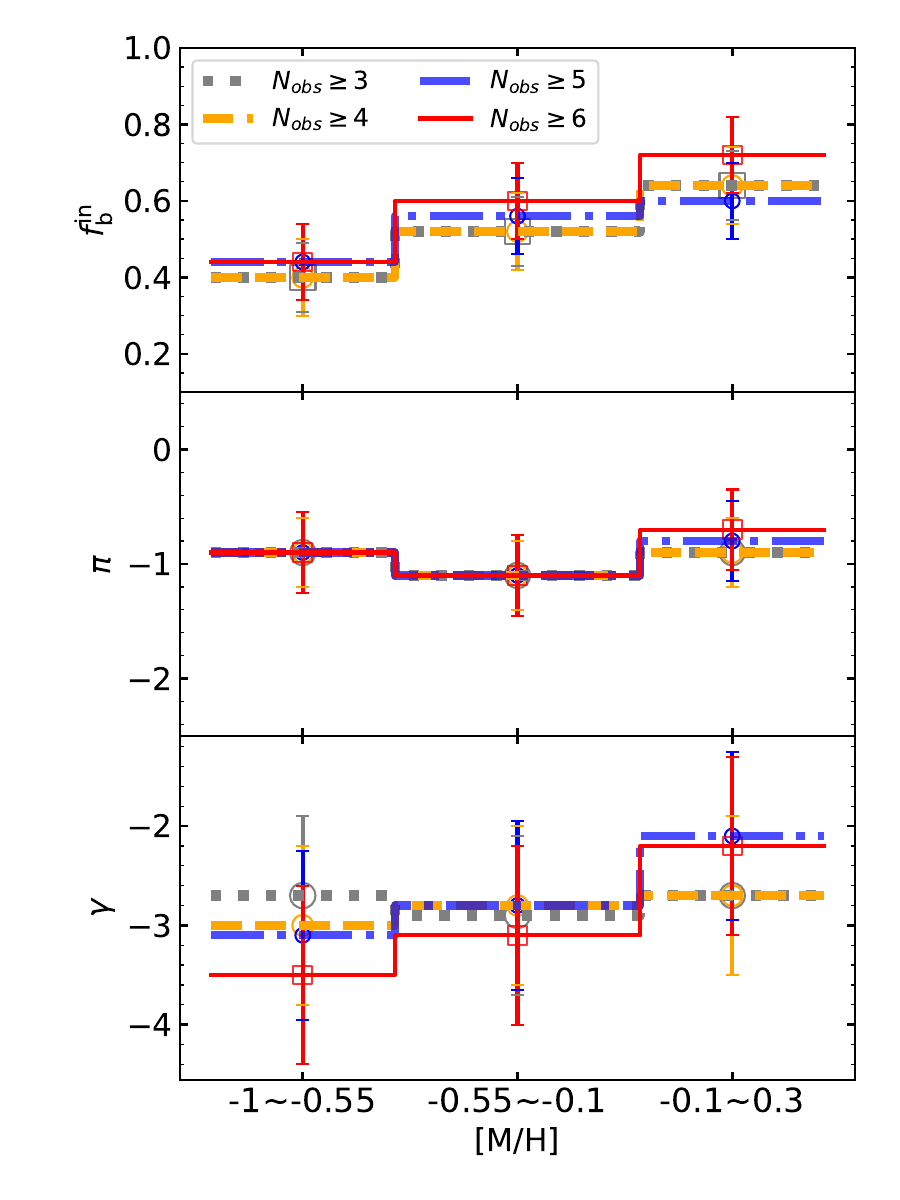}
                \caption{Intrinsic statistical properties of $f_{\rm b}^{\rm in}$, $\pi$, and $\gamma$ for stars classifed based on [M/H]. The blue dot-dashed and red solid lines represent the result of stars with observational frequencies greater than five and six, while the gray and yellow dashed lines represent frequencies greater than three and four.}\label{fig: resultMH}
\end{figure}

Figure~\ref{fig: resultMH} shows the results for different metallicities. 
It is clearly shown that the binary fraction $f_{\rm b}^{\rm in}$ (the upper panel) 
increases with [M/H]. 
For the sample of $n\ge 6$, with the highest $f_{\rm b}^{\rm in}$ obtained, the intrinsic binary fraction $f_{\rm b}^{\rm in}$ is $44\%\pm10\%$, $60\%\pm10\%$, and $72\%\pm10\%$ for [M/H]$<$-0.55, -0.55 $\leq$ [M/H]<-0.1, and [M/H] $\geq$ -0.1, respectively.  
The correlation between $f_{\rm b}^{\rm in}$ and [M/H] is still an open question and could be affected by several factors, such as star formation, binary evolution, or dynamical interaction of clusters, or any combination of the three \citep{2015Hettinger}.
\cite{2009Machida} suggested that a metal-poor cluster is more likely to form binaries via cloud fragmentation, and this may explain some observations with high $f_{\rm b}^{\rm in}$ for metal-poor stars. 
However, \cite{2012Korntreff} argued that many short-period systems would merge shortly after formation since the density of gas for a newly formed cluster increases due to dynamical evolution of the cluster.
Blue stragglers, formed by binary evolution or dynamical interaction of clusters, provide evidence for these processes \citep{2000Yanny,2002Latham,2010LublueS}.

Similarly, we do not find obvious trends for the $\pi$ and $\gamma$ values with metallicity [M/H]. 
For [M/H]$<$-0.55, -0.55 $\leq$ [M/H]<-0.1, and [M/H] $\geq$ -0.1 we have
$\pi=-0.9\pm0.35$, $-1.1\pm0.35$, and $-0.7\pm0.35$, respectively,
and $\gamma=-3.5\pm0.9$, $-3.1\pm0.9$, and $-2.2\pm0.9$, respectively.

\subsection{Dependence on the projection velocity of $v\sin{i}$}
\begin{figure}
        \centering
            \includegraphics[scale=0.6]{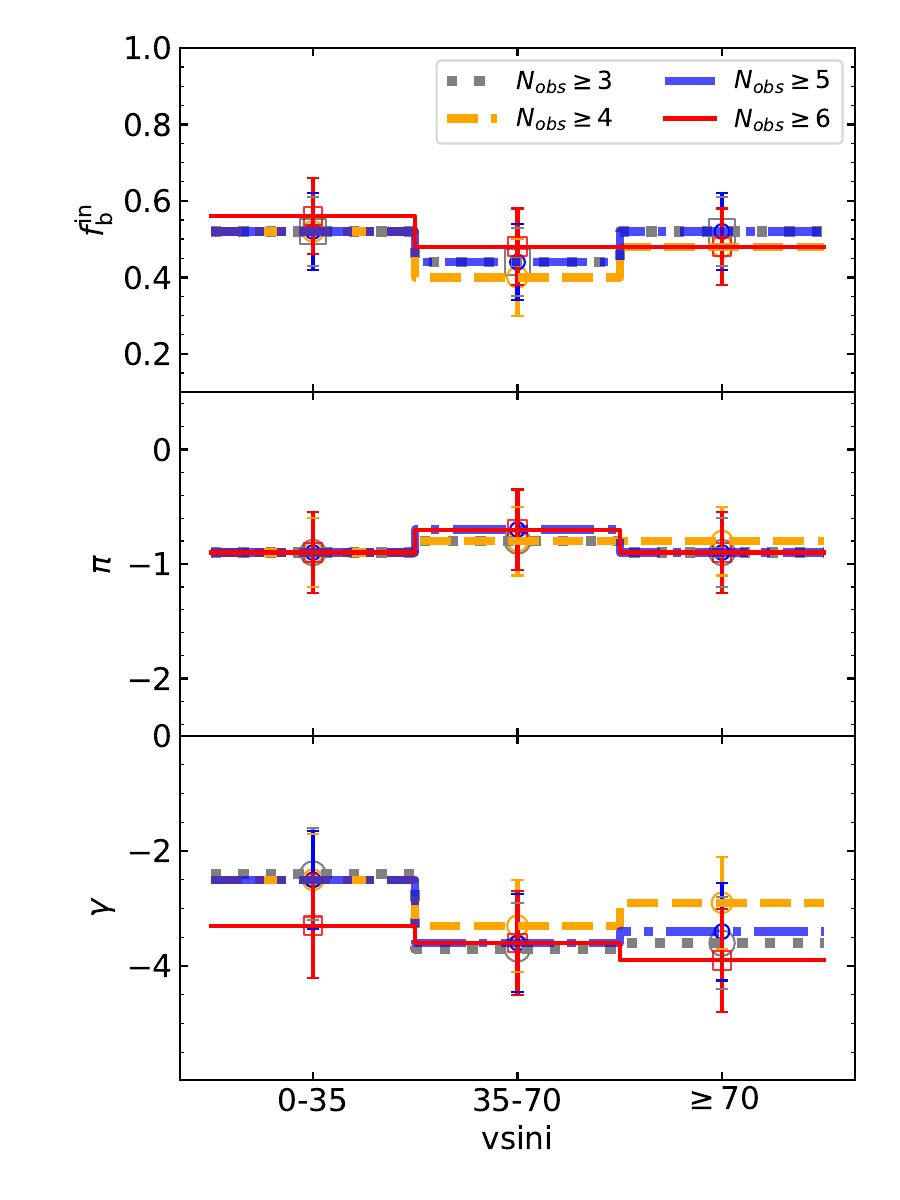}
                \caption{Intrinsic statistical properties of $f_{\rm b}^{\rm in}$, $\pi$, and $\gamma$ for stars classifed based on $v\sin{i}$. The blue dot-dashed and red solid lines represent the result of stars with observational frequencies greater than five and six, while the gray and yellow dashed lines represent frequencies greater than three and four.}\label{fig: resultvsini}
\end{figure}

\begin{figure}
        \centering
            \includegraphics[scale=0.6]{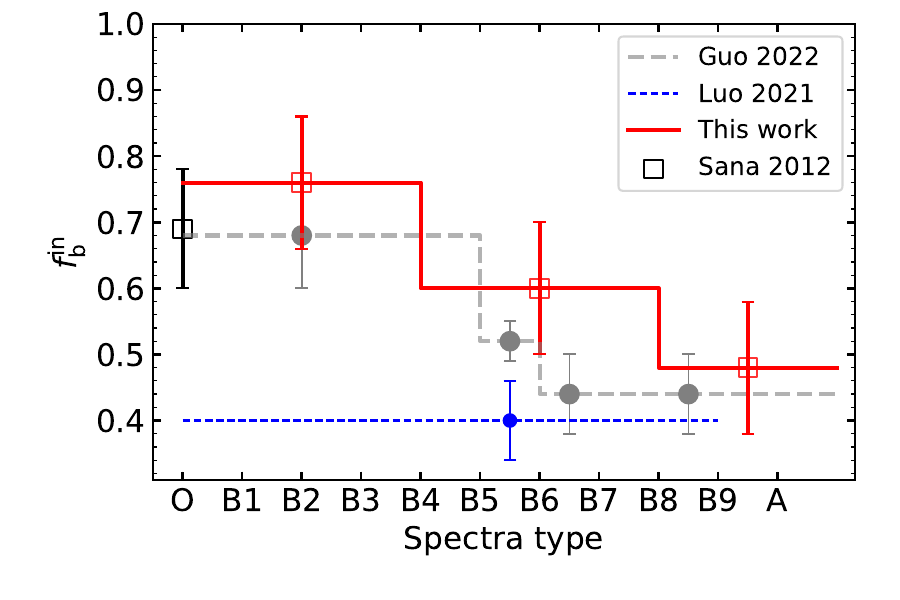}
                \caption{Intrinsic statistical properties of $f_{\rm b}^{\rm in}$ for stars with observational frequencies greater than six, classified based on $T_{\rm eff}$. 
                The red solid line represents the result of stars classified based upon $T_{\rm eff}$ in groups of B8-A, B4-B7, O-B3. The gray dashed line represents the trend of \cite{2021GYJfb} and the black square and blue symbol represent the result of \citet{2013sana} and  \citet{2021luofeng}, respectively.}\label{fig: result_T}
\end{figure}

\begin{figure}
        \centering
            \includegraphics[scale=0.6]{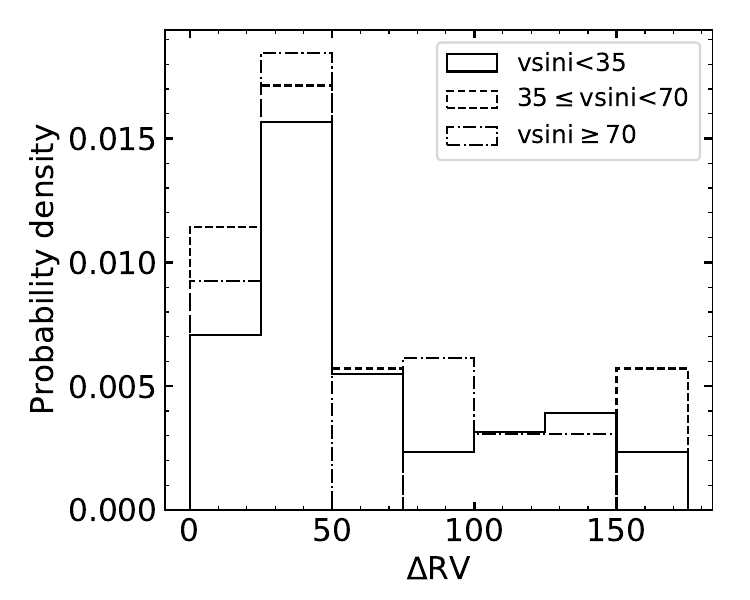}
                \caption{Distribution of $\Delta$RV in groups of $v\sin{i}$ $<$ 35 $\ {\rm km\ s^{-1}}$ (solid line), 35 $\leq$ $v\sin{i} <$70$\ {\rm km\ s^{-1}}$ (dashed line), and the stars with $v\sin{i}$ larger than 70 $\ {\rm km\ s^{-1}}$ (dot-dashed line).}\label{fig: disRV}
\end{figure}

\begin{figure}
        \centering
            \includegraphics[scale=0.7]{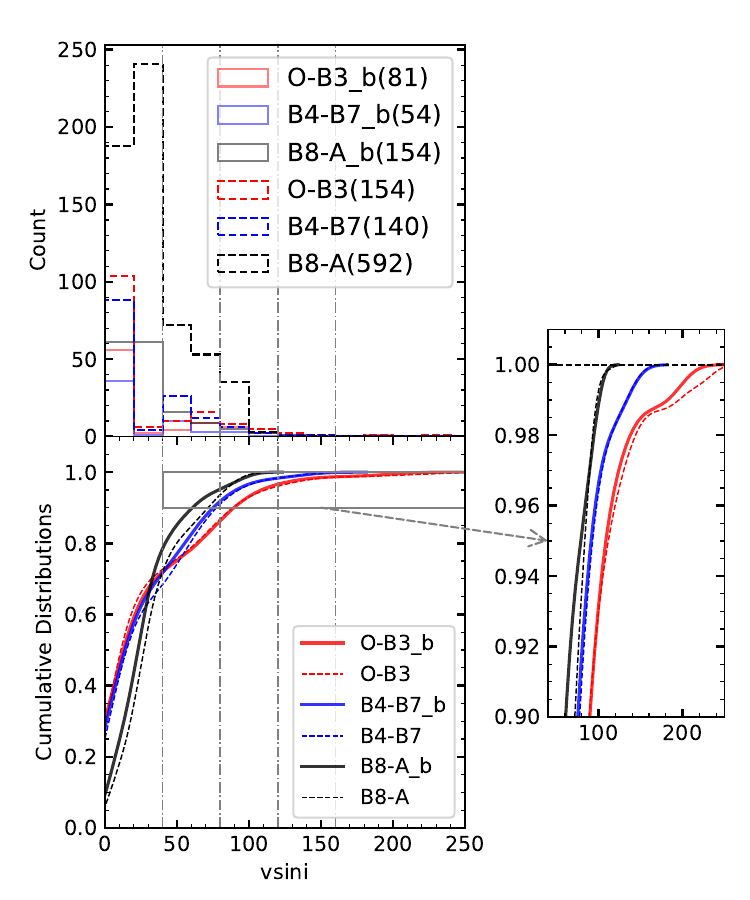}
                \caption{Comparison of the $v\sin{i}$ histograms (top panel) and cumulative distributions (bottom panel) based on different groups of $T_\mathrm{eff}$. The solid and dashed lines represent all the stars and binary samples, respectively. The red, blue, and black lines represent the samples in O-B3, B4-B7, and B8-A groups, respectively.}\label{fig: disTvsini}
\end{figure}

\begin{figure}
        \centering
            \includegraphics[scale=0.7]{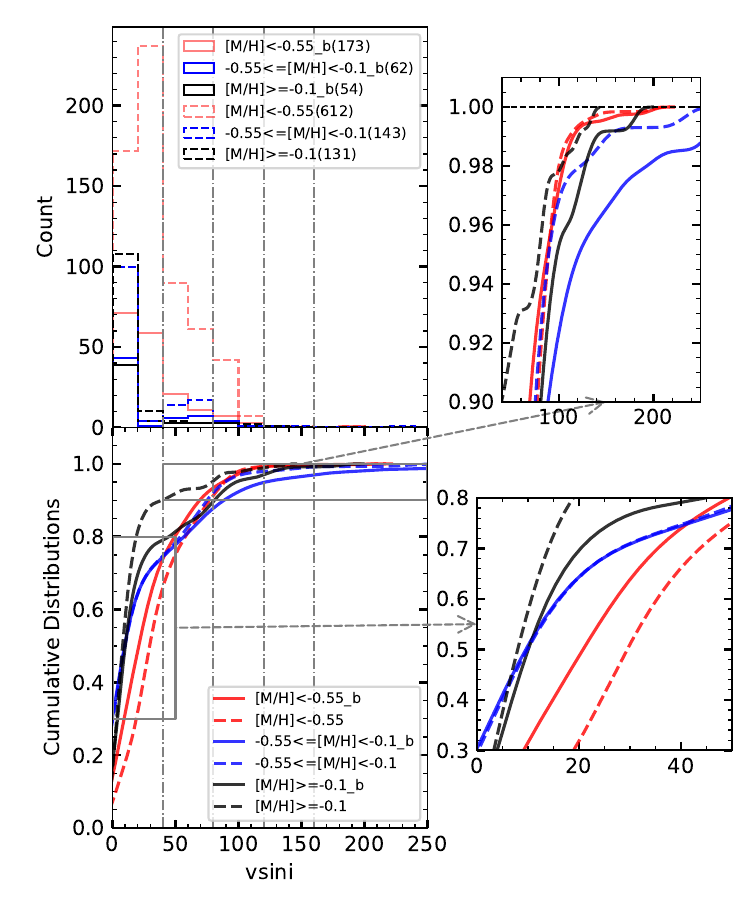}
                \caption{Comparison of the $v\sin{i}$ histograms (top panel) and cumulative distributions (bottom panel) based on different groups of [M/H]. The solid and dashed lines represent all the stars and binary samples, respectively. The red, blue, and black lines represent the samples in [M/H]$<$-0.55, -0.55 $\leq$ [M/H]<-0.1, and [M/H] $\geq$ -0.1 groups, respectively.}\label{fig: disMHvsini}
\end{figure}

The results for the dependence of statistical properties on the projection velocity are shown in Fig.~\ref{fig: resultvsini}.
There is no evidence showing the correlation between the statistical parameters and $v\sin{i}$. 
The intrinsic binary fraction $f_{\rm b}^{\rm in}$ is around 50\% for all cases, as shown in the figure.
For the sample of $n\ge 6$, $f_{\rm b}=56\%\pm10\%$, $48\%\pm10\%$, and $48\%\pm10\%$; $\pi=-0.9\pm0.35$, $-0.7\pm0.35$, and $-0.9\pm0.35$; $\gamma=-3.3\pm0.9$, $-3.6\pm0.9$, and $-3.9\pm0.9$ for $v\sin{i}$ $<$35 $\ {\rm km\ s^{-1}}$, 35 $\leq$ $v\sin{i}$<70$\ {\rm km\ s^{-1}}$, and $v\sin{i}$ $\geq$70 $\ {\rm km\ s^{-1}}$, respectively.

The absorption lines of fast rotators ($v\sin{i}$ over 200$\ {\rm km\ s^{-1}}$) 
become very flat in medium-resolution spectra, 
which can only be detected by high-resolution spectra.
These objects were therefore filtered out when we selected early-type stars 
based on equivalent widths of H and HeI lines \citep{2021GYJfb}.
The number of the fast rotators is very small 
and does not have a significant impact on the results.
On the other hand, the flat absorption lines would cause large RV errors, 
leading to an overestimate of $f_{\rm b}^{\rm in}$
(since Equation (1) could be satisfied more easily).
In any case, the stars in groups of $v\sin{i}$ $<$ 70 $\ {\rm km\ s^{-1}}$ are not affected.
For stars with $v\sin{i}$ $>$ 70 $\ {\rm km\ s^{-1}}$, 
we display the $\Delta$RV distribution in Fig.~\ref{fig: disRV} 
in comparison with the other two groups. 
We found that the three distributions are similar, 
indicating that the measurement of $v\sin{i}$ in our sample 
is not significantly affected by the RV measurement. 
So the dependence of $f_{\rm b}^{\rm in}$ on $v\sin{i}$ would not change due to these errors.

Although the values of $v\sin{i}$ estimated by the machine learning are systematically lower than those derived from high-resolution spectra (see Fig. 8 of \citealt{2021GYJslam}), 
the result that $f_{\rm b}^{\rm in}$ has no correlation with $v\sin{i}$ does not change.
This is consistent with previous studies; for example, 
Fig. 1 at the website\footnote{\url{https://aa.oma.be/stellar_rotation}} \citep{2005vsini} shows that O, B, and A stars rotate faster than the stars later than A, 
while there is no obvious dependence between spectral type and $v\sin{i}$ among O, B, and A stars. 
Our study is for O, B, and A stars, and is in close agreement. 

Since the value of $v\sin{i}$ may give hints to star formation and binary evolution, we further compare the frequencies of binary population and the whole sample on the value of $v\sin{i}$. 
Figure~\ref{fig: disTvsini} shows the results based on the spectral type of the stars and Fig.~\ref{fig: disMHvsini} based on [M/H]. 
In both figures, the upper panel shows the distribution of $v\sin{i}$ of the sample and the bottom panel is for the cumulative distribution (CDF).  
We see obvious differences in the CDFs between the binary populations and the whole samples.

In Fig.~\ref{fig: disTvsini}, for stars of O-B3 type (in red), the CDF of the whole sample, in comparison to that of the binary population, is larger when $v\sin{i}<40\ {\rm km\ s^{-1}}$ and gradually becomes smaller after that. 
This variation indicates that 
the likely single stars in the sample have a low-$v{\rm sin}i$ group 
and a high-$v{\rm sin}i$ group, while the binaries have a relatively flat distribution of $v{\rm sin}i$. 
This is possibly related to stellar evolution and binary interaction. 
The low-$v{\rm sin}i$ group of the likely single stars is the result of strong wind and magnetic braking of massive single stars \citep{2005Matt,2008Ekstr,2012Ekstr}, while the high-$v{\rm sin}i$ group is likely from the merger of binary stars \citep{2013deMink}. Binary interaction gives the binary population a relatively wide range of $v{\rm sin}i$ (or orbital periods if tidally locked).
The case of stars with of types B4-B7 (in blue) is similar, but the high-$v{\rm sin}i$ group of the likely single stars seems to have a relatively low $v{\rm sin}i$ value (in the range of 50-100$\ {\rm km\ s^{-1}}$) in comparison to that of O-B3 stars.
It is a little different for B8-A-type stars (in black). 
The CDF of the binary population is larger than that of the whole sample when $v\sin{i}<100\ {\rm km\ s^{-1}}$, and becomes smaller after that. 
The CDF of the binary population also indicates a relatively wide range of binaries. 
In particular, there are some binaries having $v{\rm sin}i$ slightly higher ($v\sin{i}>95\ {\rm km\ s^{-1}}$) than that of likely single stars.

The differences of the CDFs for various metallicity groups are more obvious than those grouped by spectral type. 
For those with high [M/H], most stars in the whole sample have low $v\sin{i}$, while the binary population span on a wide range of $v\sin{i}$. 
For those with intermediate [M/H], 
the CDFs of the whole sample and the binary population are very close when $v\sin{i} < 40\ {\rm km\ s^{-1}}$. About 70\%\ of both samples have $v\sin{i}$ less than 40$\ {\rm km\ s^{-1}}$. 
After $v\sin{i} > \sim 40\ {\rm km\ s^{-1}}$, the CDF of the whole sample increases faster 
than that of the binary population and is close to 1.0 when $v\sin{i} = \sim 240\ {\rm km\ s^{-1}}$, 
indicating the majority of the remaining 30\% of the stars have $v\sin{i}$ in 40-240$\ {\rm km\ s^{-1}}$ in the whole sample. Similarly, the CDFs of the binary population span a very wide range of $v\sin{i}$ values. About 1.5\% of the binaries have $v\sin{i}$ that can be larger than 250$\ {\rm km\ s^{-1}}$, as shown by the CDF.
We also see that a small part of the binary population have higher values of $v\sin{i}$ in comparison to that of the whole sample for the case of low [M/H]. 

Several factors affect the values of $v\sin{i}$, 
from the fragmentation of clouds and the  stellar wind of massive stars to the binary or dynamical interaction.
It is hard to assess all the effects in quantity presently,
and it is beyond the scope of this manuscript.  
It is clear to us, as shown in Figs.~\ref{fig: disTvsini} and \ref{fig: disMHvsini}, 
that the binary population is quite evenly distributed over a wide range of $v\sin{i}$ values.
This property probably comes from binary evolution or interaction (i.e., the components of a binary are tidally locked with the orbit).

\section{Conclusion}\label{sec:Conclusion}
We collected 886 early-type stars with more than six observations from LAMOST DR8, divided the sample in three ways based upon effective temperature, metallicity [M/H], and projection velocity $v\sin{i}$, and investigated the statistical properties of the samples.

We first used the variations of radial velocities of each sample 
to identify the binaries in the sample and to obtain the observed binary fraction. 
Based on the Monte Carlo simulations in statistics, we corrected observational biases and estimated the intrinsic statistical properties of the binary fraction, the distributions of orbital period, and the mass ratio. 

Our study shows that the intrinsic binary fraction increases with increasing $T_\mathrm{eff}$,
consistent with what is found  in the  literature. 
We also find that the binary fraction is positively correlated with metallicity in our sample,
consistent with \cite{1983Carney} and \cite{2015Hettinger}, but opposite to what is found in \cite{2018Tianzhijia} and \cite{2019liuchaosmokinggun}.
We do not find correlations between binary fraction and projection velocity $v\sin{i}$. However,
it seems that the binary population is relatively evenly distributed over a wide range of $v\sin{i}$ values, while the whole sample shows that most of the stars are concentrated at low values of $v\sin{i}$, and
at high values of $v\sin{i}$ in some cases. Binary evolution may partly account for this. 

We examined the uncertainties in our method induced by the sample size and observation frequency systematically in the paper and found that the uncertainties of the statistical properties decrease with increasing sample size and the observation frequency. 
For the real sample of the LAMOST DR8, 
the uncertainty of $f_{\rm b}^{\rm in}$ is similar (around 0.1) to that of the samples 
with observation frequency $n \ge 3, 4, 5,$ and 6.
For the cases of $n \ge 5$ and 6, we obtain the binary fraction of $76\%\pm10\%$, $60\%\pm10\%$, and $48\%\pm10\%$ for stars of  O-B3, B4-B7, and B8-A type, respectively. 
The binary fraction becomes smaller ($f_{\rm b}^{\rm in}=48\%$) for the samples of $n \ge 3$ and 4 
since it has a low possibility to obtain the maximum RV difference of a binary, resulting in fewer binaries to be verified in these cases, in comparison to that of relatively high observation frequencies.

We do not find obvious correlations between the orbital period distribution $f(P)$
(or the mass ratio distribution $f(q)$) and effective temperature $T_{\rm eff}$, 
and between $f(P)$ (or $f(q)$) and metallicity [M/H], 
likely due to the short observational cadence. 
For the sample with $n \ge 6$, we have $\pi=-0.9\pm0.35$, $-0.9\pm0.35$, and $-0.9\pm0.35$, and $\gamma=-1.9\pm0.9$, $-1.1\pm0.9$, and $-2\pm0.9$ respectively for stars of  O-B3, B4-B7, and B8-A types.

\begin{acknowledgements} 
This work is supported by the Natural Science Foundation of China (Nos.\ 2021YFA1600403/1, 12125303, 11733008, 12090040/3, 12103085) and by the China Manned Space Project of No. CMS-CSST-2021-A10. C.L.\ acknowledges National Key R$\&$D Program of China No.\ 2019YFA0405500 and the NSFC with grant No.\ 11835057.

Guoshoujing Telescope (the Large Sky Area Multi-Object Fiber Spectroscopic Telescope LAMOST) is a National Major Scientific Project built by the Chinese Academy of Sciences. Funding for the project has been provided by the National Development and Reform Commission. LAMOST is operated and managed by the National Astronomical Observatories, Chinese Academy of Sciences.

This work is also supported by the Key Research Program of Frontier Sciences, CAS, Grant No.\ QYZDY-SSW-SLH007.
\end{acknowledgements}

\bibliographystyle{aa} 
\bibliography{paper} 
\end{document}